\documentclass[prl,twocolumn,twoside,preprintnumbers,superscriptaddress,nofootinbib,letterpaper]{revtex4-1}

\usepackage{amsmath,slashed}
\usepackage{graphicx,graphics,color}
\usepackage{dcolumn}
\usepackage[hyperfootnotes=true,colorlinks=true,citecolor=blue,linkcolor=blue,urlcolor=blue]{hyperref}
\usepackage{xspace}
\usepackage{enumerate}
\usepackage{varwidth}

\usepackage{physics}
\usepackage{amssymb}
\usepackage{mathtools}
\usepackage{dsfont}
\usepackage{multirow}

\newcommand{\mpi}{M_\pi}

\newcommand{\meta}{M_\eta}
\newcommand{\metap}{M_{\eta'}}
\newcommand{\beq}{\begin{equation}}
\newcommand{\eeq}{\end{equation}}

\newcommand{\Order}{\mathcal{O}}

\newcommand{\GeV}{\,\text{GeV}}
\newcommand{\MeV}{\,\text{MeV}}

\newcommand{\F}{\mathcal{F}}

\newcommand{\Mr}{M_\rho}
\newcommand{\Meff}{M_\text{eff}}
\newcommand{\Ma}{M_{a_1}}
\newcommand{\Mf}{M_{f_1}}
\newcommand{\Mfp}{M_{f_1'}}
\newcommand{\Mphi}{M_\phi}

\allowdisplaybreaks[1]

\begin{document}

\preprint{PSI-PR-25-04, UWThPh 2025-8, ZU-TH 10/25}

\title{Improved evaluation of the electroweak contribution to muon $\boldsymbol{g-2}$}

\author{Martin Hoferichter}
\affiliation{Albert Einstein Center for Fundamental Physics, Institute for Theoretical Physics, University of Bern, Sidlerstrasse 5, 3012 Bern, Switzerland}
\author{Jan L\"udtke}
\affiliation{Faculty of Physics, University of Vienna, Boltzmanngasse 5, 1090 Vienna, Austria}
\author{Luca Naterop}
\affiliation{Physik-Institut, Universit\"at Z\"urich, Winterthurerstrasse 190, 8057 Z\"urich, Switzerland}
\affiliation{PSI Center for Neutron and Muon Sciences, 5232 Villigen PSI, Switzerland}
\author{Massimiliano Procura}
\affiliation{Faculty of Physics, University of Vienna, Boltzmanngasse 5, 1090 Vienna, Austria}
\author{Peter Stoffer}
\affiliation{Physik-Institut, Universit\"at Z\"urich, Winterthurerstrasse 190, 8057 Z\"urich, Switzerland}
\affiliation{PSI Center for Neutron and Muon Sciences, 5232 Villigen PSI, Switzerland}

\begin{abstract} 
A precise evaluation of the electroweak contribution to the anomalous magnetic moment of the muon requires control over all aspects of the Standard Model, ranging from Higgs physics, over multi-loop computations for bosonic and (heavy-)fermion diagrams, to non-perturbative effects in the presence of light quarks. Currently, the dominant uncertainties arise from such hadronic effects in the vector--vector--axial-vector three-point function,  an improved understanding of which has recently emerged in the context of hadronic light-by-light scattering. Profiting from these developments as well as new perturbative and non-perturbative input for the charm contribution,
    we obtain $a_\mu^\text{EW}=154.4(4)\times 10^{-11}$.
\end{abstract}

\maketitle

\emph{Introduction}---The anomalous magnetic moment of the muon $a_\mu=(g-2)_\mu/2$ has been measured to a precision of $0.19\,\text{ppm}$~\cite{Muong-2:2023cdq,Muong-2:2024hpx},
\beq
a_\mu^\text{exp}=116\,592\,059(22)\times 10^{-11},
\eeq
projected to improve to
 $\Delta a_\mu^\text{exp}\simeq 13\times 10^{-11}$ with the upcoming release of the final result from the Fermilab experiment, possibly even improving upon the original design goal~\cite{Muong-2:2015xgu}. To confront the latest experimental determination with an up-to-date Standard-Model prediction, a community-wide effort~\cite{Colangelo:2022jxc} is currently ongoing to revise the 2020 review~\cite{Aoyama:2020ynm} according to the latest developments~\cite{WP2}. In this context, the limiting factor concerns the unclear situation regarding hadronic vacuum polarization, in which case tensions among data-driven evaluations ~\cite{Davier:2017zfy,Keshavarzi:2018mgv,Colangelo:2018mtw,Hoferichter:2019gzf,Davier:2019can,Keshavarzi:2019abf,Hoid:2020xjs,Crivellin:2020zul,Keshavarzi:2020bfy,Malaescu:2020zuc,Colangelo:2020lcg,Stamen:2022uqh,Colangelo:2022vok,Colangelo:2022prz,Hoferichter:2023sli,Hoferichter:2023bjm,Stoffer:2023gba,Davier:2023fpl,CMD-3:2023alj,CMD-3:2023rfe,Leplumey:2025kvv,Hoferichter:2025lcz} and with lattice QCD~\cite{Borsanyi:2020mff,Ce:2022kxy,ExtendedTwistedMass:2022jpw,FermilabLatticeHPQCD:2023jof,RBC:2023pvn,Boccaletti:2024guq,Blum:2024drk,Djukanovic:2024cmq,Bazavov:2024eou} need to be resolved to match the experimental precision, e.g., regarding the role of radiative corrections~\cite{Campanario:2019mjh,Ignatov:2022iou,Colangelo:2022lzg,Monnard:2021pvm,Abbiendi:2022liz,BaBar:2023xiy,Aliberti:2024fpq}. For hadronic light-by-light (HLbL) scattering~\cite{Melnikov:2003xd,Masjuan:2017tvw,Colangelo:2017qdm,Colangelo:2017fiz,Hoferichter:2018dmo,Hoferichter:2018kwz,Gerardin:2019vio,Bijnens:2019ghy,Colangelo:2019lpu,Colangelo:2019uex,Pauk:2014rta,Danilkin:2016hnh,Jegerlehner:2017gek,Knecht:2018sci,Eichmann:2019bqf,Roig:2019reh} (as well as higher-order hadronic effects~\cite{Calmet:1976kd,Kurz:2014wya,Colangelo:2014qya,Hoferichter:2021wyj}), the situation is already much better, with recent work both in lattice QCD~\cite{Blum:2019ugy,Chao:2021tvp,Chao:2022xzg,Blum:2023vlm,Fodor:2024jyn} and using phenomenological approaches suggesting a viable path towards the required precision. The latter includes work on exclusive hadronic states~\cite{Hoferichter:2020lap,Zanke:2021wiq,Danilkin:2021icn,Stamen:2022uqh,Ludtke:2023hvz,Hoferichter:2023tgp,Hoferichter:2024fsj,Ludtke:2024ase,Deineka:2024mzt,Holz:2024lom,Holz:2024diw} in a dispersive approach to HLbL scattering~\cite{Colangelo:2014dfa,Colangelo:2014pva,Colangelo:2015ama,Hoferichter:2013ama}, higher-order short-distance constraints~\cite{Bijnens:2020xnl,Bijnens:2021jqo,Bijnens:2022itw,Bijnens:2024jgh}, and the combination with hadronic descriptions~\cite{Leutgeb:2019gbz,Cappiello:2019hwh,Knecht:2020xyr,Masjuan:2020jsf,Ludtke:2020moa,Colangelo:2021nkr,Leutgeb:2021mpu,Colangelo:2024xfh,Leutgeb:2024rfs,Mager:2025pvz}.  In particular, there is now a complete dispersive evaluation of the HLbL contribution at a precision of $\Delta a_\mu^\text{HLbL}=8\times 10^{-11}$~\cite{Hoferichter:2024vbu,Hoferichter:2024bae,Hoferichter:2025fea}, including a detailed study of the matching to constraints from perturbative QCD (pQCD) and the operator product expansion (OPE), which relates the HLbL tensor in parts of the parameter space to the vector--vector--axial-vector ($VVA$) correlator~\cite{Melnikov:2003xd,Vainshtein:2002nv,Knecht:2003xy}.

\begin{figure}[t]
  \centering
\includegraphics[width=0.3\linewidth]{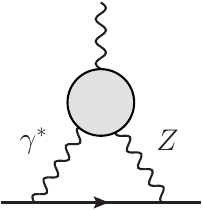}\qquad
\includegraphics[width=0.3\linewidth]{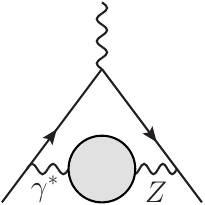}
\caption{EW diagrams afflicted by hadronic effects, including the $VVA$ correlator (left) and the $\gamma$--$Z$ two-point function (right), both indicated by gray blobs. The two diagrams are sensitive to the axial-vector and vector components of the $Z$ boson, respectively.}
\label{fig:diagrams}
\end{figure}

 In this Letter, we aim to explore the consequences of these latter developments for the electroweak (EW) contribution $a_\mu^\text{EW}$. That is, while for QED the dominant uncertainty arises from six-loop effects enhanced by powers of $\log(m_\mu/m_e)$~\cite{Aoyama:2012wk,Aoyama:2019ryr}, the entire class of EW contributions, defined as all diagrams involving $Z$, $W$, or $h$, includes loops of light quarks, and their non-perturbative manifestation proves to be the dominant source of uncertainty~\cite{Czarnecki:2002nt,Gnendiger:2013pva}, see Fig.~\ref{fig:diagrams}. In view of the fact that the evaluation of $a_\mu^\text{EW}$ in Refs.~\cite{Aoyama:2020ynm,Gnendiger:2013pva} still relies on the non-perturbative estimates from Ref.~\cite{Czarnecki:2002nt}, as well as recent work that has led to an improved understanding of the $VVA$ correlator~\cite{Ludtke:2024ase}, it is timely that these estimates be reconsidered.

 First, even for the heavy quarks new information has become available, thanks to the $\alpha_s$ corrections calculated in Ref.~\cite{Melnikov:2006qb}. Next, we can build on the first-generation analysis from Ref.~\cite{Ludtke:2024ase}, which we adapt to the strange quark, before combining the result with a new estimate of the charm contribution including the consideration of non-perturbative effects due to $\eta_c$ poles. Finally, also the evaluation of non-perturbative effects in $\gamma$--$Z$ mixing can be made more robust, incorporating recent results from lattice QCD~\cite{Ce:2022eix,Erler:2024lds}. Throughout, we follow the conventions of Ref.~\cite{Gnendiger:2013pva}, i.e., the parameters are determined via $G_F$ as measured in muon decay, $\alpha$, and $m_t$, in terms of which $M_W$ is predicted, and the on-shell scheme for the weak mixing angle $s_W^2\equiv \sin^2\theta_W=1-M_W^2/M_Z^2$ is employed. Changes in the input parameters compared to Ref.~\cite{Aoyama:2020ynm} are small, so that for the one-loop result~\cite{Jackiw:1972jz,Bars:1972pe,Altarelli:1972nc,Bardeen:1972vi,Fujikawa:1972fe} and the two-loop bosonic~\cite{Czarnecki:1995sz,Heinemeyer:2004yq,Gribouk:2005ee} and Higgs~\cite{Czarnecki:1995wq,Gnendiger:2013pva} contributions the numbers remain almost identical~\cite{Stockinger,WP2}.

\emph{Perturbative corrections}---The left diagram in Fig.~\ref{fig:diagrams} can be expressed in terms of the $VVA$ correlator, decomposed into a longitudinal ($L$) and transversal ($T$) contribution.
After performing the Wick rotation, one obtains~\cite{Jegerlehner:2017gek}
\begin{align}
\label{amu_VVA}
 a_\mu^{VVA}&=\frac{\alpha}{\pi}\frac{G_F}{24\pi^2\sqrt{2}}\int_0^\infty dQ^2\, Q^2\sum_{i=L,T}c_i(Q^2) w_i(Q^2),\notag\\
 c_L(Q^2)&=\bigg(1-\frac{Q^2}{2m_\mu^2}\bigg)W_\mu+\frac{Q^2}{2m_\mu^2},\quad W_\mu=\sqrt{1+\frac{4m_\mu^2}{Q^2}},\notag\\
 c_T(Q^2)&=\bigg[\bigg(2+\frac{Q^2}{2m_\mu^2}\bigg)W_\mu-\bigg(3+\frac{Q^2}{2m_\mu^2}\bigg)\bigg]\frac{M_Z^2}{M_Z^2+Q^2},
\end{align}
where the normalization conventions can be most easily read off from the perturbative one-loop expressions
~\cite{Czarnecki:2002nt,Rosenberg:1962pp,Adler:1969gk,Kukhto:1992qv}
\begin{align}
 w_L^{1\text{-loop}}(Q^2)&=
 2w_T^{1\text{-loop}}(Q^2)\notag\\
 &=4T_f^3 N_f Q_f^2\int_0^1 dx\frac{x(1-x)}{x(1-x)Q^2+m_f^2}\notag\\
 &=\frac{4T_f^3 N_f Q_f^2}{Q^2}\bigg(1+\frac{2m_f^2}{Q^2 W_f}\log\frac{W_f-1}{W_f+1}\bigg),
\end{align}
for a fermion $f$ with mass $m_f$, third component of weak isospin $T_f^3$, charge $Q_f$, and a potential color factor $N_f$.
The $\alpha_s$ corrections for massive quarks were calculated in Ref.~\cite{Melnikov:2006qb}, and provided as a Pad\'e series in $Q^2/m_f^2$. Using these expressions, we find for the contribution from the third generation
\beq
\label{third}
a_\mu^{VVA}[t,b,\tau]=-8.12(1)\times 10^{-11},
\eeq
where the remaining perturbative error, estimated using either a fixed $\alpha_s(m_f^2)$ or one-loop running $\alpha_s(Q^2)$ in the respective $Q^2$ integrals, is amply covered by the indicated uncertainty.\footnote{Throughout, we use mass parameters from Ref.~\cite{ParticleDataGroup:2024cfk}, in particular, $m_t=172.57(29)\GeV$, $\bar m_b(m_b)=4.183(7)\GeV$, $\bar m_c(m_c)=1.2730(46)\GeV$, and convert the latter two into on-shell masses at $\Order(\alpha_s)$, consistent with the order to which we are working, yielding $m_b=4.58\GeV$, $m_c=1.48\GeV$.  The uncertainty is dominated by higher orders in $\alpha_s$, which we estimate by comparing $\alpha_s(Q^2)$ and $\alpha_s(m_f^2)$ evaluations of the $VVA$ contribution.
For the top quark, this also covers the corrections that would need to be applied
to the Monte Carlo $m_t$ parameter quoted above.}  We recall that only the combined contribution of each generation is physical, as reflected by the anomaly cancellation condition
\beq
\sum_f N_f Q_f^2 T_f^3=0,
\eeq
which ensures that the integral in Eq.~\eqref{amu_VVA} converges.

\emph{Light quark loops}---For the first generation, we use the results of the dedicated dispersive calculation of the triplet $VVA$ correlator from Ref.~\cite{Ludtke:2024ase}
\begin{align}
 a_\mu^{VVA,\,L}[u,d,e]&=-0.892(10)\times 10^{-11},\\
 a_\mu^{VVA,\,T}[u,d,e]&=-1.192(29)\times 10^{-11},\notag\\
 a_\mu^{VVA}[u,d,e]&\equiv a_\mu^{VVA,\, L+T}[u,d,e]=-2.08(3)\times 10^{-11}.\notag
\end{align}
As preparation for an improved estimate of the second generation, we first reconstruct a simplified version of this result. Following Ref.~\cite{Hoferichter:2024bae}, we find
\begin{align}
\label{wLT_ud}
 w_L^{ud}(Q^2)&=\frac{2Q^2\big[Q^2+\mpi^2+\Mr^2+\Meff^2-\kappa_\text{OPE}^{ud}\big]}{(Q^2+\mpi^2)(Q^2+\Mr^2)(Q^2+\Meff^2)}\\
 &\qquad+\frac{\mpi^2\Mr^2\Meff^2w_L^{ud}(0)}{(Q^2+\mpi^2)(Q^2+\Mr^2)(Q^2+\Meff^2)},\notag\\
 w_T^{ud}(Q^2)&=\frac{Q^2\big[Q^2+\Ma^2+\Mr^2+\Meff^2-\kappa_\text{OPE}^{ud}\big]}{(Q^2+\Ma^2)(Q^2+\Mr^2)(Q^2+\Meff^2)}\notag\\
 &\qquad +\frac{\Ma^2\Mr^2\Meff^2w_T^{ud}(0)}{(Q^2+\Ma^2)(Q^2+\Mr^2)(Q^2+\Meff^2)},\notag
\end{align}
implementing the normalizations from chiral perturbation theory~\cite{Knecht:2020xyr,Masjuan:2020jsf}
\beq
 w_L^{ud}(0)=\frac{8\pi^2(1+a_\pi)F_{\pi\gamma\gamma}F_\pi}{\mpi^2},\hspace{4pt}
 w_T^{ud}(0)=\frac{8\pi^2a_\pi F_{\pi\gamma\gamma}F_\pi}{\mpi^2},
\eeq
with slope parameter $a_\pi=31.5(9)\times 10^{-3}$~\cite{Hoferichter:2018kwz,Hoferichter:2014vra}, pion decay constant $F_\pi=92.32(10)\MeV$~\cite{ParticleDataGroup:2024cfk}, form-factor normalization $F_{\pi\gamma\gamma}=0.2754(21)\GeV^{-1}$~\cite{PrimEx-II:2020jwd},
and the OPE constraint
\begin{align}
w_L^{ud}(Q^2)\big|_\text{OPE}&=
2w_T^{ud}(Q^2)\big|_\text{OPE}\notag\\
&=\frac{2}{Q^2}\bigg[1-\frac{\kappa_\text{OPE}^{ud}}{Q^2}+\Order\Big(\big(Q^2\big)^{-2}\Big)\bigg],
\end{align}
with subleading coefficient
\beq
\kappa_\text{OPE}^{ud}
=\frac{8\pi^2\hat m\hat X}{3e}\simeq 0.012\GeV^2,
\eeq
where we used the tensor coefficients
$X_u=40.7(1.3)\MeV$, $X_d=39.4(1.4)\MeV$, $X_s=53.0(7.2)\MeV$~\cite{Bali:2012jv,Bijnens:2019ghy,Ludtke:2024ase},
the quark masses $\hat m=3.49(7)\MeV$, $m_s=93.5(8)\MeV$~\cite{ParticleDataGroup:2024cfk,BMW:2010ucx,RBC:2012cbl,FermilabLattice:2018est,Lytle:2018evc,Bruno:2019vup,ExtendedTwistedMass:2021gbo}, and neglected isospin-breaking effects beyond $\hat m=(m_u+m_d)/2$, $\hat X=(X_u+X_d)/2$.
As observed in Ref.~\cite{Hoferichter:2024bae}, the dispersive result of Ref.~\cite{Ludtke:2024ase} is almost perfectly reproduced for effective masses $\Meff=1.5\GeV$ and $\Meff=1.0\GeV$ for the longitudinal and transversal components, respectively, producing $a_\mu^{VVA}[u,d,e]=-2.07(6)\times 10^{-11}$, where the uncertainty indicates the sensitivity to varying $\Meff\in[1.0,2.0]\GeV$.

For the generalization to the strange-quark case, the most complicated aspect concerns the $\eta$--$\eta'$ and $f_1$--$f_1'$ mixings. For the normalizations we have
\begin{align}
w_L^{s}(0)&=-4\pi^2\sum_{P=\eta,\eta'}\frac{(1+b_PM_P^2)F_{P\gamma\gamma}F_P^s}{M_P^2},\notag\\
w_T^{s}(0)&=-4\pi^2\sum_{P=\eta,\eta'}b_PF_{P\gamma\gamma}F_P^s,
\end{align}
with slope parameters $b_\eta = 1.833(41)\GeV^{-2}$, $b_{\eta'} = 1.493(32)\GeV^{-2}$~\cite{Holz:2024diw,Holz:2022hwz}, normalizations $F_{\eta\gamma\gamma}=0.2736(48)\GeV^{-1}$, $F_{\eta'\gamma\gamma}=0.3437(55)\GeV^{-1}$~\cite{ParticleDataGroup:2024cfk}, and strangeness decay constants $F_P^s$. The latter can be reconstructed from the singlet and octet decay constants and mixing angles~\cite{Hoferichter:2022mna,Holz:2024diw} (consistent with Refs.~\cite{Escribano:2015yup,Bali:2021qem,Ottnad:2025zxq}), yielding
\beq
F_\eta^s=-111.7(4.4)\MeV,\qquad F_{\eta'}^s=140.8(4.8)\MeV,
\eeq
and similarly, the relative weight of $f_1$, $f_1'$ can be estimated from light-cone sum rules for the strangeness couplings $F_{A}^s$~\cite{Yang:2007zt} and the experimental normalizations $\F_2^{A}(0,0)$~\cite{Achard:2001uu,Achard:2007hm}. Defining
\begin{align}
\xi_P&=\frac{F_{P\gamma\gamma}F_P^s}{\sum_{P'=\eta,\eta'}F_{P'\gamma\gamma}F_{P'}^s},\notag\\
  \xi_A&=\frac{\frac{F_{A}^s}{M_{A}}\F_2^{A}(0,0)}{\sum_{A'=f_1,f_1'}\frac{F_{A'}^s}{M_{A'}}\F_2^{A'}(0,0)},
\end{align}
these weights become
\beq
\label{weights}
\xi_\eta\simeq -1.7,\quad \xi_{\eta'}\simeq 2.7,\quad
\xi_{f_1}\simeq -2.0,\quad \xi_{f_1'}\simeq 3.0,
\eeq
which together with the
OPE constraint
\begin{align}
w_L^{s}(Q^2)\big|_\text{OPE}&=
2w_T^{s}(Q^2)\big|_\text{OPE}\notag\\
&=-\frac{2}{3Q^2}\bigg[1-\frac{\kappa_\text{OPE}^{s}}{Q^2}+\Order\Big(\big(Q^2\big)^{-2}\Big)\bigg],\notag\\
\kappa_\text{OPE}^{s}
&=\frac{8\pi^2m_sX_s}{3e}\simeq 0.43\GeV^2,
\end{align}
leads to the following generalization of Eq.~\eqref{wLT_ud}:
\begin{widetext}
\begin{align}
\label{wLT_s}
 w_L^{s}(Q^2)&=\Bigg\{-\frac{2Q^2\big[Q^2+\xi_\eta\meta^2+\xi_{\eta'}\metap^2+\Mphi^2+\Meff^2-\kappa_\text{OPE}^{s}\big]}{3(Q^2+\Mphi^2)(Q^2+\Meff^2)}
 +\frac{M_{\eta\eta'}^2\Mphi^2\Meff^2w_L^{s}(0)}{(Q^2+\Mphi^2)(Q^2+\Meff^2)}\Bigg\}\sum_{P=\eta,\eta'}\frac{\xi_P}{Q^2+M_P^2},\\
 w_T^{s}(Q^2)&=\Bigg \{-\frac{Q^2\big[Q^2+\xi_{f_1}\Mf^2+\xi_{f_1'}\Mfp^2+\Mphi^2+\Meff^2-\kappa_\text{OPE}^{s}\big]}{3(Q^2+\Mphi^2)(Q^2+\Meff^2)}
 +\frac{M_{f_1f_1'}^2\Mphi^2\Meff^2w_T^{s}(0)}{(Q^2+\Mphi^2)(Q^2+\Meff^2)}\Bigg\}\sum_{A=f_1,f_1'}\frac{\xi_A}{Q^2+M_A^2},\notag
\end{align}
\end{widetext}
where
\beq
M_{\eta\eta'}^2=\frac{M_\eta^2 M_{\eta'}^2}{\xi_\eta M_{\eta'}^2+\xi_{\eta'}M_{\eta}^2},\qquad
M_{f_1f_1'}^2=\frac{\Mf^2 \Mfp^2}{\xi_{f_1} \Mfp^2+\xi_{f_1'}\Mf^2}.
\eeq

As reference point, we first evaluate the second generation with a leading-order (LO) pQCD loop for charm
\begin{align}
 a_\mu^{VVA,\,L}[c,s,\mu]\big|_\text{LO pQCD}&=-2.92(2)\times 10^{-11},\notag\\
 a_\mu^{VVA,\,T}[c,s,\mu]\big|_\text{LO pQCD}&=-1.70(5)\times 10^{-11},\notag\\
 a_\mu^{VVA}[c,s,\mu]\big|_\text{LO pQCD}&=-4.62(5)\times 10^{-11},
\end{align}
where the indicated errors refer to only the surprisingly small sensitivity to $\Meff$. In fact, by far the largest uncertainty arises from the relative weight of the $\eta$ and $\eta'$ contributions, determined in Eq.~\eqref{weights} via the two-photon couplings and the strangeness decay constants. This could be improved by a dedicated dispersive analysis; here, we assign the variation compared to the weights $\xi_\eta=-1$, $\xi_{\eta'}=2$~\cite{Czarnecki:2002nt}, corresponding to the limit in which $\eta$--$\eta'$ mixing is neglected in $F_{P\gamma\gamma}$ and $F_P^s$, as an additional uncertainty, which yields
\beq
a_\mu^{VVA}[c,s,\mu]\big|_\text{LO pQCD}=-4.62(14)\times 10^{-11}.
\eeq

 \begin{figure}[t]
  \centering
\includegraphics[width=\linewidth]{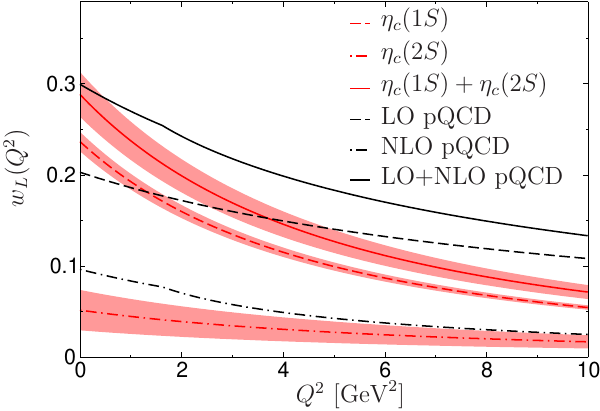}
\caption{$w_L(Q^2)$ for the charm loop. The black lines refer to pQCD (dashed: LO, dot-dashed: NLO, solid: sum), while the $\eta_c$ estimate according to Eq.~\eqref{wL_etac} is represented by the red lines (dashed: $\eta_c(1S)$, dot-dashed: $\eta_c(2S)$, solid: sum). The bands are propagated from the uncertainties in $F_P$ and $F_{P\gamma\gamma}$.}
\label{fig:wL_etac}
\end{figure}

\emph{Charm loop}---For the charm contribution, it is clear that perturbative corrections will be more sizable than in the context of Eq.~\eqref{third}, but also non-perturbative effects could play a role for small $Q^2$. Turning first to the $\alpha_s$ corrections from Ref.~\cite{Melnikov:2006qb}, with a running $\alpha_s$ kept constant below $Q_0=m_c$, one obtains the shifts
\begin{align}
\label{NLO}
 \Delta a_\mu^{VVA,\,L}[c]\big|_{\alpha_s}&=0.33(17)\times 10^{-11},\notag\\
\Delta a_\mu^{VVA,\,T}[c]\big|_{\alpha_s}&=0.14(7)\times 10^{-11},\notag\\
\Delta a_\mu^{VVA}[c]\big|_{\alpha_s}&=0.48(24)\times 10^{-11},
\end{align}
where the uncertainties indicate the variation observed when keeping $\alpha_s(Q_0)$ fixed (for both $L$ and $T$ in the upward direction).
However, this evaluation assumes the validity of pQCD in the entire integration region, while for small $Q^2$ a hadronic description should be used instead. The most important intermediate states are
 the $\eta_c(1S)$, $\eta_c(2S)$ poles, suggesting the estimate
\beq
\label{wL_etac}
 w_L^{\eta_c}(Q^2)=\sum_{P=\eta_c(1S),\,\eta_c(2S)}\frac{8\pi^2 F_{P\gamma\gamma}F_{P}}{M_{P}^2+Q^2}\frac{M_V^2}{M_V^2+Q^2},
 \eeq
 where the vector mass is set to $M_V=M_{J/\psi}$ and $M_V=M_{\psi(2S)}$, respectively.\footnote{This is motivated by the fact that the decay $\psi(2S)\to\gamma \eta_c(2S)$ has been observed~\cite{BES:2012uhz}, while for $\eta_c(2S)\to\gamma J/\psi$ only limits are available~\cite{BESIII:2017tsq}. However, the difference between the scales is small and only has negligible impact on the analysis.}
 The $\eta_c(1S)$ decay constant was calculated in lattice QCD (including quenched QED) in Ref.~\cite{Hatton:2020qhk}, $F_{\eta_c(1S)}=398.1(1.0)\MeV$, which together with the two-photon coupling $F_{\eta_c(1S)\gamma\gamma}=0.067(3)\GeV^{-1}$~\cite{ParticleDataGroup:2024cfk} then defines the simplest estimate of non-perturbative effects.\footnote{The recent lattice-QCD calculation from Ref.~\cite{Colquhoun:2023zbc} yields $F_{\eta_c(1S)\gamma\gamma}=0.0781(4)\GeV^{-1}$, translating into an increased value of $w_L^{\eta_c}(Q^2)$ (see also Ref.~\cite{Meng:2021ecs}). However, the resulting curve would still be consistent with the pQCD prediction.}
 To include, in addition, the $\eta_c(2S)$, we use $F_{\eta_c(2S)\gamma\gamma}=0.032(11)\GeV^{-1}$~\cite{ParticleDataGroup:2024cfk} and $F_{\eta_c(2S)}=271(69)\MeV$~\cite{Chung:2020zqc}, whose result $F_{\eta_c(1S)}=385(94)\MeV$ agrees well with Ref.~\cite{Hatton:2020qhk}.

 As shown in Fig.~\ref{fig:wL_etac}, this model matches the sum of LO and next-to-leading-order (NLO) pQCD almost perfectly in the limit $Q^2\to 0$, so that, in practice, non-perturbative corrections prove small and the pQCD description can be used in the entire domain. More rigorously, we can match the hadronic and pQCD description at a sufficiently high scale at which pQCD starts to apply,  incurring negligible changes compared to Eq.~\eqref{NLO}. In the spirit of Ref.~\cite{Ludtke:2024ase}, a smooth matching
 in the intermediate range could also be achieved by modeling the effects of even heavier intermediate states by effective poles, again leading to results compatible within uncertainties.
 One could further try to extract non-perturbative input for $w_T(Q^2)$ from the radiative axial-vector decay $\chi_{c1}\to\gamma J/\psi$,  following Refs.~\cite{Zanke:2021wiq,Hoferichter:2023tgp}, but the required model assumptions would be too severe to allow for an improved estimate.
 Overall, we therefore obtain
 \begin{align}
 \Delta a_\mu^{VVA}[c]\big|_{\alpha_s+\eta_c}&=0.48(24)\times 10^{-11},\notag\\
 a_\mu^{VVA}[c,s,\mu]&=-4.14(28)\times 10^{-11}.
 \end{align}

\emph{$\gamma\text{--}Z$ mixing}---The second class of non-perturbative contributions, represented by the right diagram in Fig.~\ref{fig:diagrams}, is conventionally included
 in the fermionic remainder, once Higgs and $VVA$ contributions have been separated. Writing this $\gamma$--$Z$-mixing term in the form
\beq
a_\mu^{\gamma Z}=-\frac{G_F m_\mu^2}{8\pi^2\sqrt{2}}\frac{\alpha}{\pi}\times \frac{4}{3}(1-4s_W^2)\times 8\pi^2\bar\Pi^{\gamma Z}(-M_Z^2),
\eeq
with the correlator and running couplings defined in the conventions
\begin{align}
\bar\Pi^{\gamma Z}(q^2)&=-\frac{s_W^2}{4\pi\alpha}\Delta_\text{had}s_W^2(q^2),\notag\\
\Delta_\text{had}s_W^2(q^2)&=\Delta\alpha_\text{had}(q^2)-\Delta\alpha_{2,\,\text{had}}(q^2),\notag\\
\alpha(q^2)&=\frac{\alpha(0)}{1-\Delta \alpha_\text{had}(q^2)-\Delta\alpha_\text{lep}(q^2)},
\end{align}
one has the perturbative expression
\beq
8\pi^2\bar\Pi^{\gamma Z}(-M_Z^2)=2\sum_{q=u,d,s,c,b} \Big[T_q^3 Q_q-2Q_q^2 s_W^2\Big]\log\frac{M_Z}{m_q},
\eeq
which needs to be amended due to non-perturbative effects. In the evaluation of Refs.~\cite{Aoyama:2020ynm,Czarnecki:2002nt,Gnendiger:2013pva} the replacement
$8\pi^2\bar\Pi^{\gamma Z}(-M_Z^2)\to 6.88(50)$ is used, which goes back to Ref.~\cite{Marciano:1993jd} (with input from Refs.~\cite{Marciano:1983ss,Jegerlehner:1985gq,Jegerlehner:1990uiq}). The more recent data-driven update from Ref.~\cite{Jegerlehner:2017gek}, also using the methodology for the flavor separation from Ref.~\cite{Jegerlehner:1985gq}, finds $5.87(4)$ instead. To assess the $SU(3)$ corrections in the flavor decomposition in a more robust manner, we consider the lattice-QCD input from Ref.~\cite{Ce:2022eix} (see Ref.~\cite{Erler:2024lds} for the time-like evolution to $q^2=M_Z^2$), which amounts to writing
\begin{align}
 8\pi^2\bar\Pi^{\gamma Z}(-M_Z^2)&=\frac{\pi}{\alpha}\big(1-2s_W^2\big)\Delta\alpha_\text{had}^{(5)}(-M_Z^2)\notag\\
 &-\frac{4\pi^2}{3\sqrt{3}}\bar\Pi^{08}(-M_Z^2)\notag\\
 &-\frac{2\pi^2}{3}\sum_{q=c,b}Q_q\bar\Pi^{qq}(-M_Z^2),
\end{align}
neglecting small isospin-breaking and heavy-quark disconnected contributions.
This formulation shows explicitly
that the by far numerically dominant contribution can be retrieved from $e^+e^-\to\text{hadrons}$ cross sections, while $\bar\Pi^{08}$ parameterizes the amount of $SU(3)$ breaking. Using
$\Delta\alpha_\text{had}(-M_Z^2)=0.02756(10)$~\cite{Davier:2019can,Keshavarzi:2019abf,Jegerlehner:2019lxt}, the first term gives $6.56(3)$, consistent with the lattice+pQCD evaluation from Ref.~\cite{Ce:2022eix}, $6.59(4)$.
The subsequent terms lead to a reduction of the central value. Taking $\bar\Pi^{08}(-M_Z^2)=7.0(2)\times 10^{-3}$~\cite{Ce:2022eix} (extrapolated from $q^2=-7\GeV^2$ to $q^2=-M_Z^2$ with a Pad\'e ansatz), $\bar\Pi^{08}$ induces a shift by about $-0.05$, while the pQCD result for $q=c,b$~\cite{Kallen:1955fb,Chetyrkin:1996cf} adds another $-0.52(4)$ (with the uncertainty covering the difference between LO and NLO in $\alpha_s$). Taking everything together, we find a combined value
$8\pi^2\bar\Pi^{\gamma Z}(-M_Z^2)=6.0(1)$,\footnote{We also considered the results for the pQCD corrections given in Ref.~\cite{Erler:2023hyi}, both for $\Delta \alpha_\text{had}$ and $\bar \Pi^{qq}$, leading to a result for $\bar\Pi^{\gamma Z}(-M_Z^2)$ consistent within uncertainties.} leading to an uncertainty in $\Delta a_\mu^{\gamma Z}$ that is negligible compared to the uncertainty from terms suppressed by higher powers in $1-4s_W^2$  or $M_Z^2/m_t^2$, see Ref.~\cite{Gnendiger:2013pva}, whose error estimate we take over. Updating particle masses to Ref.~\cite{ParticleDataGroup:2024cfk} then leads to the value for the fermionic remainder given in Table~\ref{tab:summary}.

\begin{table}[t]
	\renewcommand{\arraystretch}{1.3}
	\centering
\begin{tabular}{lrr}
\toprule
& Ref.~\cite{Aoyama:2020ynm} & This work\\\colrule
One-loop & $194.79(1)$ & $194.79(1)^*$\\
Two-loop, bosonic & $-19.96(1)$ & $-19.96(0)^*$\\
Two-loop, Higgs & $-1.51(1)$ & $-1.50(0)^{*}$\\
Two-loop, $VVA$, $[u,d,e]$ & $-2.28(20)$ & $-2.08(3)$\\
Two-loop, $VVA$, $[c,s,\mu]$ & $-4.63(30)$ & $-4.14(28)$\\
Two-loop, $VVA$, $[t,b,\tau]$ & $-8.21(10)$ & $-8.12(1)$\\
Two-loop, fermionic (rest) & $-4.64(10)$ & $-4.58(10)$\\
Three-loop, NLL & $0.00(20)$ & $0.00(20)^*$\\\colrule
Total & $153.56(1.00)$ & $154.41(36)$\\
\botrule
\end{tabular}
\caption{Our evaluation of $a_\mu^\text{EW}$ in comparison to Ref.~\cite{Aoyama:2020ynm} (based on Refs.~\cite{Czarnecki:2002nt,Gnendiger:2013pva}; entries marked with an asterisk are taken over from Ref.~\cite{Aoyama:2020ynm}, updated to current values for particle masses~\cite{Stockinger,WP2}). All numbers are in units of $10^{-11}$.}
	\label{tab:summary}
\end{table}

\emph{Bottom line}---Summing everything up, we obtain the total as given in Table~\ref{tab:summary}, i.e.,
\beq
\label{total}
a_\mu^\text{EW}=154.4(4)\times 10^{-11}.
\eeq
This result is almost $1\sigma$ larger than the previous evaluation from Refs.~\cite{Aoyama:2020ynm,Czarnecki:2002nt,Gnendiger:2013pva}, with changes primarily driven by (i) dispersive evaluation of the $VVA$ correlator for the first generation ($+0.20$), (ii) $\eta$--$\eta'$ mixing in $w_L^s$ ($+0.14$), (iii) $\alpha_s$ and non-perturbative corrections for charm loop ($+0.48$), and all these effects happen to go in the same direction.

At this point, the largest uncertainties originate from the second generation, therein mainly due to charm physics, and the estimate of three-loop corrections by renormalization-group arguments~\cite{Degrassi:1998es,Czarnecki:2002nt}. In Table~\ref{tab:summary} we took over the previous estimate for next-to-leading logarithms (NLL), improvements of which could lie within reach using modern EFT technology~\cite{Naterop:2024ydo}.

\emph{Conclusions}---In this Letter we revised the evaluation of the EW contribution to the anomalous magnetic moment of the muon, building on an improved understanding of the required two- and three-point functions that has become available thanks to recent work on the $VVA$ correlator in the context of HLbL scattering, $\alpha_s$ corrections to the heavy quark loops, and the running of $\sin^2\theta_W$. In total, we find a rather sizable shift compared to previous work, see Eq.~\eqref{total} for the final result, but consistent within uncertainties and driven by the accumulation of a number of small improvements that happen to point in the same direction. While some further refinements could be envisioned, mainly in the context of the second-generation $VVA$ contribution and three-loop NLL estimates, our work provides a reassessment of the EW contribution including a state-of-the-art evaluation of non-perturbative effects.

\begin{acknowledgments}
\emph{Acknowledgments}---We thank Nora Brambilla,  Aida El-Khadra, Jens Erler, Rodolfo Ferro-Hern\'andez, Simon Holz, Simon Kuberski, and Dominik St\"ockinger for valuable discussions, as well as Dominik St\"ockinger and Hyejung St\"ockinger-Kim for providing updated values for the two-loop bosonic and Higgs contributions.
Financial support by the SNSF (Project Nos.\ TMCG-2\_213690 and PCEFP2\_194272) as well as by the FWF (DACH Grant I 3845-N27 and Doctoral Program Particles and Interactions, project no.\ W1252-N27) is gratefully acknowledged. This research was supported in part by grant NSF PHY-2309135 to the Kavli Institute for Theoretical Physics (KITP).
\end{acknowledgments}

\bibliography{amu}

\end{document}